\documentclass[10pt,aps,prb,twocolumn,showpacs,preprintnumbers,superscriptaddress,amsmath,amssymb,longbibliography]{revtex4-1}

\usepackage{graphicx} 

\usepackage[english]{babel}
\usepackage{blindtext}
\usepackage{dcolumn}
\usepackage{bm}
\usepackage{ulem}

\newcommand{\co}[2]{\ifcase #1 \or #2 \fi}

\newif\ifnote

\begin{document}

\title{3D simulations of the electrothermal and THz emission properties of Bi$_2$Sr$_2$CaCu$_2$O$_8$ intrinsic Josephson junction stacks}

\author{F.~Rudau}
\affiliation{Physikalisches Institut and Center for Quantum Science (CQ) in LISA$^+$, Universit\"{a}t T\"{u}bingen, D-72076 T\"{u}bingen, Germany}
\author{R.~Wieland}
\affiliation{Physikalisches Institut and Center for Quantum Science (CQ) in LISA$^+$, Universit\"{a}t T\"{u}bingen, D-72076 T\"{u}bingen, Germany}
\author{J.~Langer}
\affiliation{Physikalisches Institut and Center for Quantum Science (CQ) in LISA$^+$, Universit\"{a}t T\"{u}bingen, D-72076 T\"{u}bingen, Germany}
\author{X.~J.~Zhou}
\affiliation{National Institute for Materials Science, Tsukuba 3050047, Japan}
\affiliation{Research Institute of Superconductor Electronics, Nanjing University, Nanjing 210093, China}
\author{M.~Ji}
\affiliation{National Institute for Materials Science, Tsukuba 3050047, Japan}
\affiliation{Research Institute of Superconductor Electronics, Nanjing University, Nanjing 210093, China}
\author{N.~Kinev}
\affiliation{Kotel'nikov Institute of Radio Engineering and Electronics, Russia}
\author{L.~Y.~Hao}
\affiliation{National Institute for Materials Science, Tsukuba 3050047, Japan}
\affiliation{Research Institute of Superconductor Electronics, Nanjing University, Nanjing 210093, China}
\author{Y.~Huang}
\affiliation{National Institute for Materials Science, Tsukuba 3050047, Japan}
\affiliation{Research Institute of Superconductor Electronics, Nanjing University, Nanjing 210093, China}
\author{J.~Li}
\affiliation{Research Institute of Superconductor Electronics, Nanjing University, Nanjing 210093, China}
%
%
\author{P.~H.~Wu}
\affiliation{Research Institute of Superconductor Electronics, Nanjing University, Nanjing 210093, China}
\author{T.~Hatano}
\affiliation{National Institute for Materials Science, Tsukuba 3050047, Japan}
\author{V.~P.~Koshelets}
\affiliation{Kotel'nikov Institute of Radio Engineering and Electronics, Russia}
\author{H.~B.~Wang}
\affiliation{National Institute for Materials Science, Tsukuba 3050047, Japan}
\affiliation{Research Institute of Superconductor Electronics, Nanjing University, Nanjing 210093, China}
\author{D.~Koelle}
\affiliation{Physikalisches Institut and Center for Quantum Science (CQ) in LISA$^+$, Universit\"{a}t T\"{u}bingen, D-72076 T\"{u}bingen, Germany}
\author{R.~Kleiner}
\affiliation{Physikalisches Institut and Center for Quantum Science (CQ) in LISA$^+$, Universit\"{a}t T\"{u}bingen, D-72076 T\"{u}bingen, Germany}
\date{\today}

\begin{abstract}
We used 2D coupled sine-Gordon equations combined with 3D heat diffusion equations to numerically investigate the thermal and electromagnetic properties of a $250 \times 70\,\mu\mathrm{m}^2$ intrinsic Josephson junction stack. The 700 junctions are grouped to 20 segments; we assume that in a segment all junctions behave identically. At large input power a hot spot forms in the stack. Resonant electromagnetic modes, oscillating either along the length ((0, $n$) modes) or the width (($m$, 0) modes) of the stack or having a more complex structure, can be excited both with and without a hot spot. At fixed bath temperature and bias current several cavity modes can coexist in the absence of a magnetic field. The (1, 0) mode, considered to be the most favorable mode for THz emission, can be stabilized by applying a small magnetic field along the length of the stack. A strong field-induced enhancement of the emission power is also found in experiment, for an applied field around 5.9\,mT. 
\end{abstract}

\pacs{74.50.+r, 74.72.-h, 85.25.Cp}


\maketitle

\section{Introduction}
\label{sec:intro}
Stacks of intrinsic Josephson junctions (IJJs) in the high-temperature superconductor Bi$_2$Sr$_2$CaCu$_2$O$_8$ (BSCCO) emit coherent radiation at THz frequencies\cite{Ozyuzer07}.  The emitted frequency $f_\mathrm{e}$ follows the Josephson relation $f_\mathrm{e} = V_\mathrm{J}/\Phi_0$, where $\Phi_0$ is the flux quantum ($\Phi_0^{-1} = 483.6\,\mathrm{GHz/mV}$) and $V_\mathrm{J}$ is the voltage across a single junction. In BSCCO superconductivity is restricted to $d_\mathrm{s} = 0.3\,\mathrm{nm}$ thick CuO$_2$ sheets, separated by 
barrier layers to form an $s = 1.5\,\mathrm{nm}$ thick IJJ\cite{Kleiner92}. In Ref.~\onlinecite{Ozyuzer07}, stacks of $\sim$700 IJJs, with a length $L_\mathrm{s} \sim$300\,$\mu$m, and a width $W_\mathrm{s}$ of some 10\,$\mu$m have been realized as mesas on top of BSCCO single crystals. These mesas emitted radiation between 0.35 and 0.85\,THz, with an integrated output power of $\sim$ 1\,$\mu$W. The emission frequency scaled as $W_\mathrm{s}^{-1}$, indicating that cavity modes, oscillating along the width of the stack, are responsible for synchronization. 
THz radiation from IJJ stacks became a hot topic both in experiment 
\cite{Wang09a, Minami09, Kurter09,Guenon10, Kurter10, Wang10a,Tsujimoto10,Benseman11,Yuan12,Li12,Kakeya12,Tsujimoto12,Tsujimoto12a,
Kashiwagi12, An13,Benseman13,Benseman13a,Sekimoto13,Turkoglu13,Minami14,Watanabe14,Ji14,Tsujimoto14,Watanabe15, Zhou15a,Zhou15b, Hao15,Gross15,Kakeya15,Kashiwagi15a, Kashiwagi15b, Benseman15, Tsujimoto16}
and theory
\cite{Bulaevskii07, Koshelev08,Lin08,Krasnov09,Klemm09,Tachiki09,Pedersen09,Hu09,Koyama09,Krasnov10,Koshelev10,Lin10b,Katterwe10,Yurgens11,Koyama11,Krasnov11,Yurgens11b,Lin11b,Gross12, Asai12,Asai12b,Zhang12,Lin12,Grib12,Gross13,Liu13,Asai13,Asai14,Grib14,Lin14, Rudau15}; for a recent review, see Ref. \onlinecite{Welp13}. 

IJJ stacks, containing 500 -- 2000 junctions, have been patterned as mesas but also as bare IJJ stacks contacted by Au layers (GBG structures)\cite{An13,Ji14,Kashiwagi12,Sekimoto13} and as all-superconducting structures\cite{Yuan12}. Emission frequencies range from 0.3 to 2.4\,THz. For the best stacks, an emission power $P_\mathrm{e}$ in the range of tens of $\mu$W has been achieved\cite{An13,Benseman13a,Sekimoto13,Kashiwagi15a, Kashiwagi15b}, and arrays of mesas showed emission with $P_\mathrm{e}$ up to 0.61\,mW \cite{Benseman13a}. 
The physics of the huge IJJ stacks is affected by Joule heating \cite{Ozyuzer07, Wang09a,Kurter09,Guenon10,Yurgens11,Yurgens11b,Benseman13,Wang10a,Kakeya12, Asai12,Gross12, Minami14,Asai14, Gross15,Watanabe15}. For sufficiently low bias currents, the temperature rises only slightly to values above the bath temperature $T_{\rm bath}$ and the voltage $V$ across the stack increases with increasing bias current $I$. With increasing $I$ and input power the current-voltage characteristics (IVCs) start to back-bend and, at some bias current in the back-bending region, a hot spot forms in the stack \cite{Wang09a, Wang10a,Guenon10,Kakeya12,Benseman13,Turkoglu13,Minami14, Watanabe14,Tsujimoto14,Kitamura14,Watanabe15,Benseman15}, creating a region heated to temperatures above the critical temperature $T_\mathrm{c}$. Similar effects also occur in other systems \cite{Gurevich87,Spenke36b}. 
The THz emission properties of the IJJ stacks are affected by the hot spot. For example, it has been found that the linewidth of radiation is much narrower in the high-bias regime than at low bias\cite{Li12, Gross13}. 
Other properties such as the emission frequency seem to be basically independent of the hot spot position, leading to some debate as to whether the hot spot is helpful for radiation or just coexists with the radiating regions\cite{Sekimoto13,Minami14, Watanabe14}. In fact, recent results showed that there is a strong interaction\cite{Zhou15b}. Further, cooling has been improved by sandwiching the stacks between substrates with high thermal conductivity. In first attempts maximum emission frequencies near 1.05\,THz were obtained\cite{Ji14,Kitamura14}. This value was recently improved to 
2.4 THz for disk-shaped stacks \cite{Kashiwagi15b}. 
In terms of modeling, many calculations of electrodynamics have been based on a homogeneous temperature distribution, while calculations of the thermal properties were based on solving the heat diffusion equations in the absence of Josephson currents\cite{Yurgens11,Yurgens11b,Gross12}. Some attempts have been made to combine both electrodynamics and thermodynamics, either by using arrays of point-like IJJs\cite{Grib12,Grib14, Gross13} or by incorporating temperature-induced effects into an effective model describing the whole stack as a single ``giant'' junction\cite{Asai12,Asai13,Asai14}. Ref. \onlinecite{Rudau15} modelled the combined thermal and electromagnetic properties of BSCCO stacks via one-dimensional coupled sine-Gordon equations for a $N$ = 700 junction stack where the IJJs were grouped into $M$ segments\cite{Rudau15}.
 
\section{Model}
\label{sec:model}
\begin{figure}[tb]
\includegraphics[width=\columnwidth]{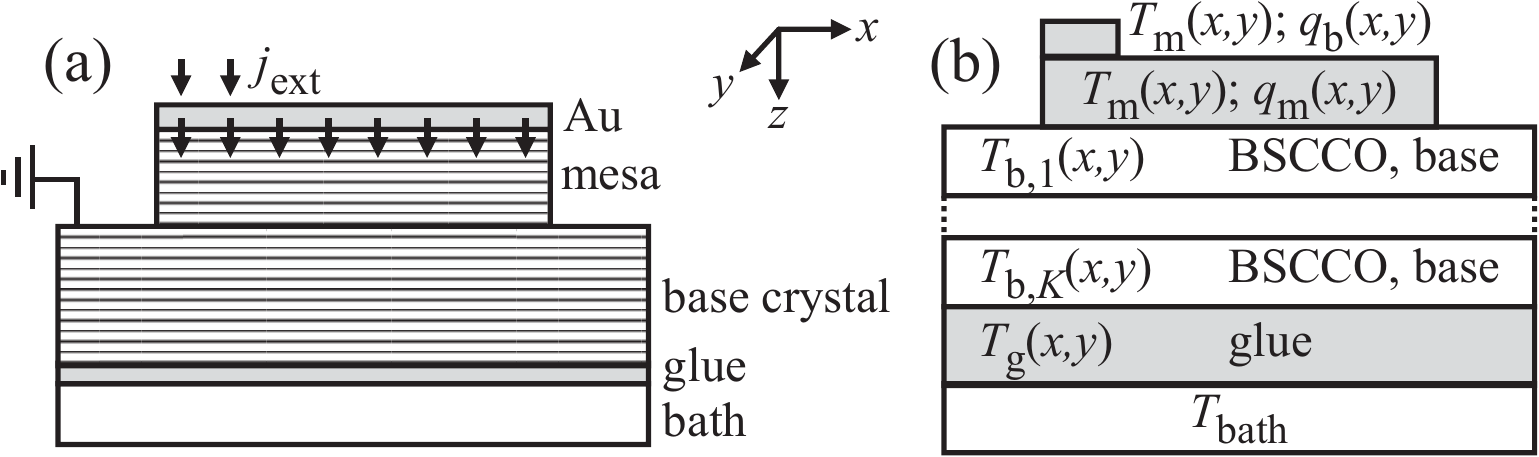}
\caption{Geometry used for modeling. Sketch of mesa and electric current flow in $z$ direction (a). Geometry for thermal description (b), where $q_{\rm m}$ and $q_{\rm b}$, respectively, denote the Joule power density produced in the mesa and by the bias lead. The temperatures of the various layers are indicated.
}
\label{fig:geometry}
\end{figure}

The model introduced here extends the 2D approach of Ref.~\onlinecite{Rudau15} to 3D, enabling us to model IJJ stacks realistically. We first give a brief outline of the features which go beyond Ref.~\onlinecite{Rudau15}. We consider a mesa consisting of $N = 700$ IJJs, cf. Fig.~\ref{fig:geometry}(a). The mesa has a length $L_{\rm s}$ = 250\,$\mu$m along $x$ and a width $W_{\rm s}$ = 70\,$\mu$m along $y$. It is covered by a gold layer and centered on a 30\,$\mu$m thick base crystal of length $L_{\rm b} = 2L_{\rm s}$ and width $W_{\rm b} = 2W_{\rm s}$. The base crystal is mounted by a 20\,$\mu$m thick glue layer to a sample holder, kept at $T_{\rm bath}$. A bias current $I$ is injected via a bond wire into the Au layer and leaves the mesa into the base crystal.
The model contains a variety of parameters (in-plane and out-of-plane resistivities, Josephson critical current density, Cooper pair density, thermal conductances, etc.) which depend on temperature. We assume that these parameters are spatially constant for spatially \textit{constant} $T$. For an inhomogeneous temperature distribution in the stack they vary in space through their dependence on the local temperature $T(x,y,z)$, which is found by self-consistently solving the thermal equations (requiring Joule heat dissipation as an input from the electric circuit) and the electrical equations (requiring the temperature distribution in the mesa, as determined from the thermal circuit).

For the thermal description, cf. Fig.~\ref{fig:geometry}(b), we assume that the mesa plus the contacting Au layer and the bond wire have a temperature $T_{\rm m}(x,y)$ which is constant along $z$ but can vary along $x$ and $y$. The BSCCO base crystal is split into $K$ segments, the $k$th segment being at a temperature $T_{{\rm b},k}(x,y)$. 
For this geometry we solve the heat diffusion equation
\begin{align}
\label{eq:heat_diffusion}
    c\dot T = \nabla(\kappa\nabla T) + q_\mathrm{m} + q_\mathrm{b},
\end{align}
%
with the specific heat capacity $c$, the (anisotropic) thermal conductivity $\kappa$ and the power densities $q_\mathrm{m}$  and $q_\mathrm{b}$ for heat generation in the mesa and the bond wire, respectively. 
For high enough $q_\mathrm{b}$, the hot spot is controllably located near the wire position. 

For the electric circuit we group the $N$ IJJs in the mesa to $M$ segments, each containing $G = N/M$ IJJs, assumed to have identical properties. 
The bond wire injects an electric current density $j_{\rm{ext}}$ to the Au layer which we assume to have a low enough resistance to freely distribute the current before it enters the IJJ stack in $z$ direction with a density $j_{z,\mathrm{Au}}$ proportional to the local BSCCO conductance $\sigma_{c} (x,y) = \rho_c^{-1}(x,y)$.
The full expression is $j_{z,\mathrm{Au}} = \left\langle j_{\rm{ext}}\right\rangle \sigma_{c}(x,y) / \left\langle\sigma_{c}\right\rangle$, the brackets denoting spatial averaging.
%
The interface stack/base crystal is treated as a ground. The $z$-axis currents consist of Josephson currents with critical current density $j_{{\rm c}}(x,y)$, (ohmic) quasiparticle currents with resistivity $\rho_c(x,y)$ and displacement currents with dielectric constant $\varepsilon$. To avoid weakly stable solutions we also add Nyquist noise created by the quasiparticle currents. The in-plane currents consist of a superconducting part, characterized by a Cooper pair density $n_{{\rm s}}(x,y)$, a quasiparticle component with resistivity $\rho_{ab}(x,y)$ and a Nyquist noise component.
For constant $T_{\rm m}(x,y)$ = 4.2\,K we index above quantities by an additional ``0'' and assume that they are constant with respect to $x$ and $y$. The temperature dependence of the various parameters is close to experimental curves and plotted in detail in Ref. \onlinecite{Rudau15}. We further use $T_{\rm c} = 85$\,K.

One obtains sine-Gordon-like equations for the Josephson phase differences $\gamma_m(x,y)$ in the $m$th segment of the IJJ stack: 
\begin{align}
\label{eq:sigo_segment}
\begin{split}
Gsd_{\rm s}\nabla(\frac{\nabla\dot{\gamma}_m}{\rho_{ab}}) +d_{\rm s}\nabla(j^{\rm N}_{{x},m+1}-j^{\rm N}_{{x},m}) + G\lambda_{\rm k}^2 \nabla(n_{\mathrm s}\nabla\gamma_m) = \\
2j_{z,m}-j_{z,m+1}-j_{z,m-1}.
\end{split}
\end{align}
Here, $m$ = $1..M$, $\nabla = (\partial/\partial x, \partial/\partial y)$ and 
$\lambda_{\rm k} = (\Phi_0 d_{\rm s}/(2\pi\mu_0j_{\rm c0}\lambda_{ab0}^2)^{1/2}$, with the in-plane London penetration depth $\lambda_{ab0}$  and the magnetic permeability $\mu_0$. Quantities $j^{\rm N}_{{x},m}$ are the in-plane noise current densities.
Time is normalized to $\Phi_0/2\pi j_{\rm c0}\rho_{c0}s$, resistivities to $\rho_{c0}$ and current densities to $j_{\rm c0}$. 
Eq. \eqref{eq:sigo_segment} neglects geometric inductances, i.e. assumes that kinetic inductances dominate (valid if $L_\mathrm{s}, W_\mathrm{s} <$ $\lambda_c$; $\lambda_c \sim$ 300\,$\mu$m is the out-of-plane penetration depth).

For the out-of-plane current densities $j_{{z},m}$ one finds
\begin{align}
\label{eq:RCSJ}
j_{{z},m} = \beta_{\rm c0} \ddot{\gamma}_m + \frac{\dot{\gamma}_m}{\rho_{{c},m}} + j_{\rm c} \sin(\gamma_m) +j^{\rm N}_{{z},m},
\end{align}
with $\beta_{\rm c0} = 2\pi j_{\rm c0}\rho_{c0}^2\varepsilon\varepsilon_0s/\Phi_0$; $\varepsilon_0$ is the vacuum permittivity and the $j^{\rm N}_{{z},m}$ are the out-of-plane noise current densities.
From the gauge invariant Josephson phase differences $\gamma_m$, as calculated from Eqs.~\eqref{eq:sigo_segment} and \eqref{eq:RCSJ}, we obtain the phase $\phi_m$ of the superconducting wave function in electrodes $m$ (the CuO$_2$ layer interfacing segments $m$ and $m + 1$) via
\begin{align}
\label{eq:phi_gamma}
    \nabla \gamma_m = \frac{2\pi s}{\Phi_0}(B_{y,m}, -B_{x,m}) + \frac{\nabla(\phi_{m+1}-\phi_m)}{G}.
\end{align}
Here, $B_{x,m}$ and $B_{y,m}$ are, respectively, the $x$ and $y$ components of the magnetic field in the $m$th segment.

The in-plane supercurrent densities in units of $j_{\mathrm{c}0}$, $\vec j^{\mathrm s}_m = (j^{\mathrm s}_{x,m}, j^{\mathrm s}_{y,m})$, in electrode $m$ are expressed as
\begin{align}
\label{eq:supercurrents}
    \vec j^{\mathrm{s}}_m = \frac{\lambda^2_\mathrm{k} n_\mathrm{s}}{d_\mathrm{s}} (\nabla\phi_m - \frac{2\pi}{\Phi_0}\vec A_m).
\end{align}
$\vec A_m = (A_{x,m}, A_{y,m})$ denotes the in-plane components of the vector potential in electrode $m$. The resistive currents $\vec j^{\mathrm r}_m = (j^{\mathrm r}_{x,m}, j^{\mathrm r}_{y,m})$ in electrode $m$ are given by
\begin{align}
\label{eq:resisitve_currents}
    \vec j^{\mathrm r}_m = \frac{s}{\rho_{ab}} \frac{\mathrm d}{\mathrm dt}(\nabla\phi_m - \frac{2\pi}{\Phi_0}\vec A_m).
\end{align}
In our calculations we assume that the $z$ components of $\operatorname{curl}\vec j^{\mathrm s}_m$ and of $\operatorname{curl} \vec j^{\mathrm r}_m$ vanish, and thus inside the superconducting layers the total magnetic field in $z$ direction is zero.

For the thermal parameters we use the same values as in Ref. \onlinecite{Rudau15}.
The bond wire with resistivity $\rho_{\rm b} = 0.02\rho_{c0}$ is assumed to be a 25\,$\mu$m wide square located at the left edge of the mesa. 
Further, $\rho_{c0} $ = 10$^3$\,$\Omega$cm, $\rho_{ab0}$ = 8\,$\mu \Omega$cm,  $j_{\rm c0} = 200$\,A/cm$^2$, $\lambda_{ab0}$ = 260\,nm and $\varepsilon = 12$.
For our geometry one obtains a critical current $I_{\rm c0}$ = 35\,mA, a $c$-axis resistance per junction $R_{c0}$ = 0.86\,$\Omega$, a characteristic voltage $V_{\rm c0} = I_{\rm c0}R_{c0}$ = 30\,mV and a characteristic frequency $f_{\rm c0} = I_{\rm c0}R_{c0}/\Phi_0$ = 14.5\,THz. 
The characteristic power density $p_{\rm c0} = j^2_{\rm c0}\rho_{c0}$ is $4\cdot10^7$\,W/cm$^3$, yielding, for a stack volume of $1.84\cdot10^{-8}$\,cm$^3$, a power $P_{\rm c0}$ of 0.74\,W. For $\lambda_\mathrm{k}$ one obtains 0.76\,$\mu$m.  
The 4.2\,K value of the in-phase mode velocity $c_1$ = $8.8\cdot10^7$\,m/s~\cite{Rudau15}.  
We keep the product $\beta_{\rm c0}G$ constant in order to (approximately) fix the 4.2\,K value of $c_1$ and use $\beta_{\rm c0}$ = 4000 for $G$ = 35 ($M$ = 20). We further divide ac electric fields and in-plane current densities by $G$ to make results only weakly dependent on $M$. For selected bias conditions the scaling has been tested using $M$ = 50.

The differential equations are discretized using 50 (9) grid points along $x$ ($y$) for the mesa and 100 (18) grid points for the base crystal
\footnote{Note that for $M$ = 20 and the relatively low number of grid points along $x$ and $y$ we cannot resolve modes that fluctuate strongly in space, like antiphase oscillations of different junctions or static triangular fluxon lattices appearing in magnetic fields on the order of a flux quantum per junction. However we are mainly interested in dynamic in-phase solutions which can be captured well with the discretization used.},
which is split into $K$ = 4 segments. 
A 5th order Runge-Kutta scheme is used to evolve these equations in time.    
%
After some initialization steps \cite{Rudau15} various quantities, partially averaged over spatial coordinates, are tracked as a function of time to produce time averages or to make Fourier transforms. 

\begin{figure}[tb]
\includegraphics[width=\columnwidth]{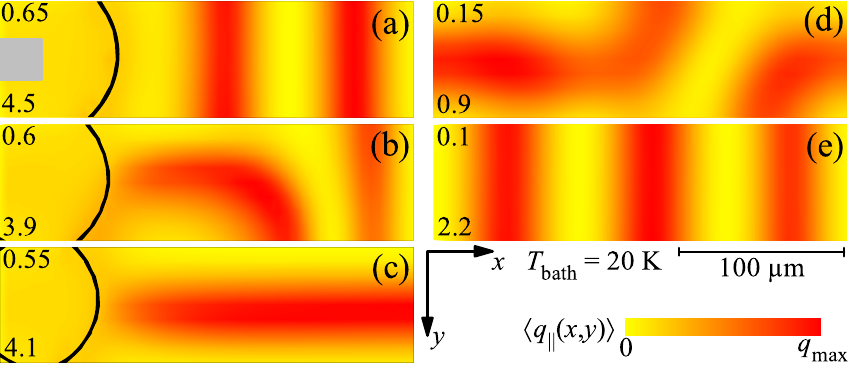}
\caption{Power density $\langle q_{\parallel}(x,y)\rangle$ in units of $10^{-5}\cdot j_\mathrm{c0}^2\rho_{c0}$ (color scale)
for five values of normalized bias current $I/I_\mathrm{c0}$ (upper left numbers); values for $q_{\rm max}$ at bottom left. The grey square in (a) indicates position of the bond wire. Regions enclosed by the black line are at $T_\mathrm{m} \geq T_\mathrm{c}$.
}
\label{fig:time_averages}
\end{figure}

\section{Results}
\label{sec:results}
Fig.~\ref{fig:time_averages} shows, for $T_{\rm bath}$ = 20\,K, averaged distributions of the power density $\langle q_{\parallel}(x,y) \rangle$ dissipated by in-plane currents for five values of $I/I_\mathrm{c0}$ = 0.65 (a) to 0.1 (e). Averaging is over time and the $z$ direction in the mesa. This type of plots, also used in Ref.~\onlinecite{Rudau15}, is useful to visualize resonance patterns, with nodes (antinodes) appearing at the minima (maxima) of $\langle q_{\parallel}(x,y) \rangle$\footnote{A perhaps more natural choice would have been to look at the time average \unexpanded{$\langle E^2_z(x,y)\rangle$} of the square of the $z$-axis electric fields. However, $E_z(x,y,z,t)$ has a large dc component and features of oscillating standing waves are only weakly visible.}. 
The left (right) graphs are at high (low) bias where a hot spot is present (absent). In (a) and (e) the modulations along $x$ are due to a cavity mode oscillating along $x$ (a $(0,n)$ mode), with $n = 2$ and 3, respectively). In (c) a cavity mode oscillating along $y$ is excited (a $(1,0)$ mode). The spatial variations in (b) and (d) have a more complicated structure which is not easy to explain by a superposition of different cavity modes. The patterns also show that ``linear thinking'' in terms of separating ac Josephson currents and resonant modes can be dangerous. Near the antinodes of the standing waves vortex/antivortex pairs oscillate back and forth, colliding at the center of the antinode\cite{Rudau15}. The collision zones should form a continuous line leaving the stack either at its edges or into the hot spot area. All patterns fulfill this requirement.

In general not all segments in the stack were synchronized. We investigated this by monitoring the dc voltages ($\propto$ Josephson oscillation frequency $f_\mathrm{J}$) $v_m$ ($m = 1\,..\,M$) across the individual segments. For example, for the modes of Figs.~\ref{fig:time_averages}(a) to (c), for the 2--3 uppermost segments $v_m$ was about 1\,\% higher than for the other (locked) segments. For the mode of Fig.~\ref{fig:time_averages}(d) only small groups of 2--5 adjacent segments were locked. For the mode of Fig.~\ref{fig:time_averages}(e) two groups of segments (1--7 and 10--20) oscillated at slightly different frequencies.

Note that $\langle q_{\parallel}(x,y)\rangle$ can have similar values for $(0,n)$ and $(1,0)$ modes, compare, eg., Figs. 2(a) and (c). We expect that both types of modes radiate.
However, for comparable values of $\langle q_\parallel(x,y)\rangle$ the \textit{emission} power of the $(0,n)$ modes, with $n>1$, will be lower, because the contributions of the oscillating (in-plane) currents to the magnetic vector potential partially cancel each other. For the $(1,0)$ mode, the in-plane currents at a given time have the same sign everywhere in the stack.
\begin{figure}[tb]
\includegraphics[width=\columnwidth]{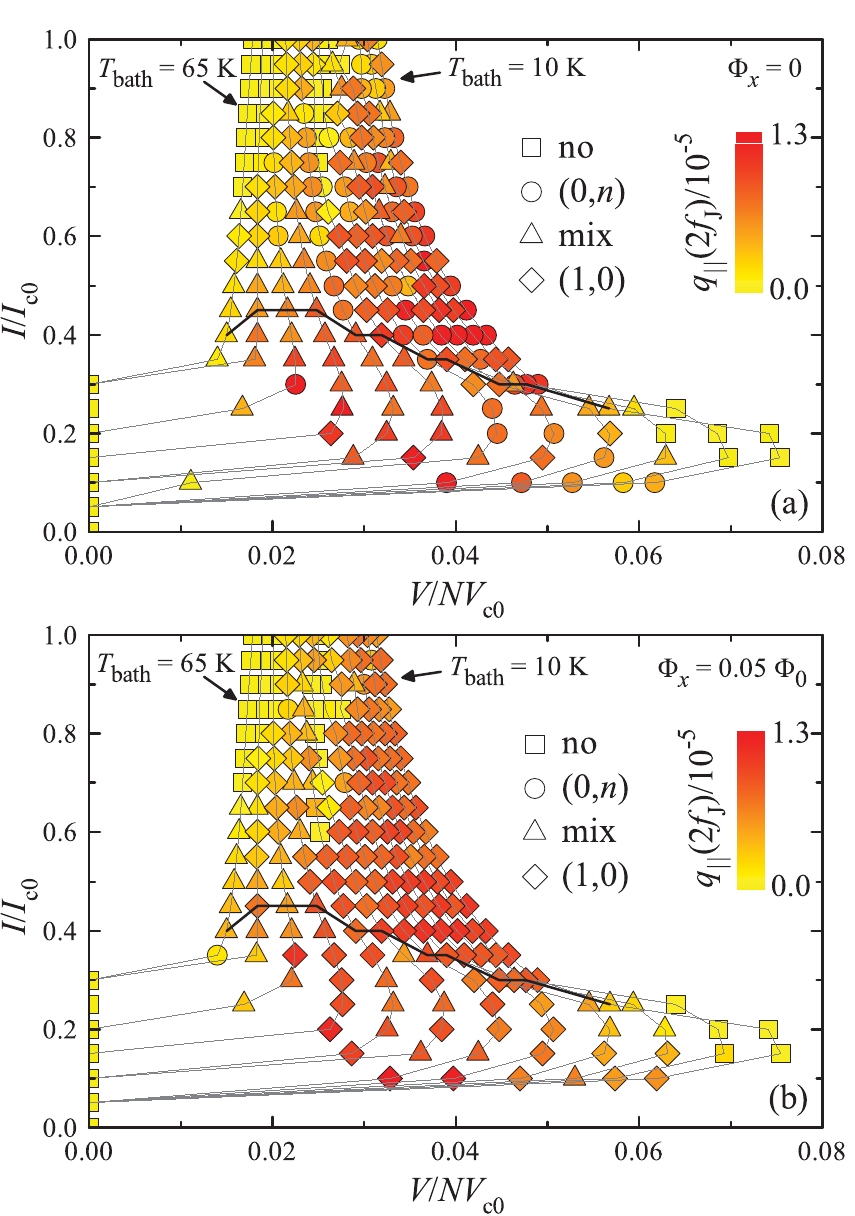}
\caption{Power density $q_{\parallel}(2f_\mathrm{J})$ in units of $10^{-5}\cdot j_\mathrm{c0}^2\rho_{c0}$ (color scale) vs. normalized bias current and voltage across the stack for an applied magnetic field along $x$ of (a) 0 and (b) 1 mT (0.05\,$\Phi_0$ per junction), applied along $x$. $T_\mathrm{bath}$ was varied from 10\,K to 65\,K in steps of 5\,K.
$(0,n)$ modes, $(1,0)$ modes, mixed resonances and nonresonant states are marked by, respectively, circles, diamonds, triangles and squares. The grey lines indicate IVCs at fixed $T_\mathrm{bath}$. For data points at or above the  black line ($T_\mathrm{c}$ line) a hotspot has formed in the stack.}
\label{fig:IVEs}
\end{figure}

Fig.~\ref{fig:IVEs}(a) shows, for zero applied magnetic field, how different modes in the stack evolve as a function of $I$ and $T_\mathrm{bath}$. We recorded 12 IVCs for $T_\mathrm{bath}$ between 10\,K and 65\,K. For each value of $I$ and $T_\mathrm{bath}$ we evaluated the type of mode by inspecting plots as in Fig.~\ref{fig:time_averages} and encoded it as the shape of the symbol in Fig.~\ref{fig:IVEs}(a). 
To have a measure of  the strength of a given mode we recorded
timetraces $q_{\parallel}(t)$ of the power generated by in-plane currents, averaged over the stack volume. After Fourier transform we extract from $q_{\parallel}(f)$ the power density $q_{\parallel}(2f_\mathrm{J})$ arising from the Josephson oscillations, appearing as a peak at twice the Josephson frequency $f_\mathrm{J}$. This quantity is plotted as the color scale for each data point. 
In Fig.~\ref{fig:IVEs}(a) there are three regions where $q_{\parallel}(2f_\mathrm{J})$ is low: (i) for $I/I_\mathrm{c0} > 0.5$ and $T_\mathrm{bath} > 55\,\mathrm{K}$, (ii) for $T_\mathrm{bath}$ around 35\,K and $I/I_\mathrm{c0} > 0.65$ and (iii) for $V/NV_\mathrm{c0} > 0.06$. In region (i) no or only a small fraction of the stack is superconducting; Josephson oscillations are absent or restricted to a small area. In region (ii) the in-plane- and out-of plane currents exhibited short-wavelength oscillations along $x$ and $y$ indicative of a mode with spatial variations shorter than our grid spacing. The spectrum of $q_{\parallel}(f)$ was broad, with no significant peaks. In region (iii), where $V$ and $f_\mathrm{J}$ were highest, all currents and fields varied smoothly, but no resonance was excited. In the presence of a hotspot (data points at or above the black line in Fig.~\ref{fig:IVEs}(a)) $q_{\parallel}(2f_\mathrm{J})$ is large in a ribbon between $V/NV_\mathrm{c0} \sim 0.025$ and 0.05. This regime extends down to $\sim$0.02 in the low-bias regime. The relative broadness of this regime may look surprizing, since resonant modes are excited, however can be understood from the facts that the mode velocities depend on temperature\cite{Rudau15} and vary significantly over the data points in Fig.~\ref{fig:IVEs}(a). Also, the quality factor of the cavity modes is low (of order 10) at elevated temperatures. Most importantly, one notes that $(0,n)$, $(1,0)$ and mixed modes vary almost randomly. Further simulations revealed that even for the same value of $I$ and $T_\mathrm{bath}$ different resonant modes can be excited.
However, it should be possible to support the $(1,0)$ mode, favored for radiation, by applying a small static magnetic field along $x$, imprinting a linear phase gradient and consequently a small gradient on the Josephson current along $y$. Fig.~\ref{fig:IVEs}(b), organized like Fig.~\ref{fig:IVEs}(a) shows the resulting data for a small field $B_x$ of 1\,mT, corresponding to a flux of $0.05\,\Phi_0$ per junction. The amplitudes of $q_{\parallel}(2f_\mathrm{J})$ are similar as in the zero field case, however, the (1,0) mode has stabilized over a wide range of bias current and bath temperature.
\section{Comparison to experiment}
\label{sec:experiment}
\begin{figure*}[tb]
\includegraphics[width=\textwidth,clip]{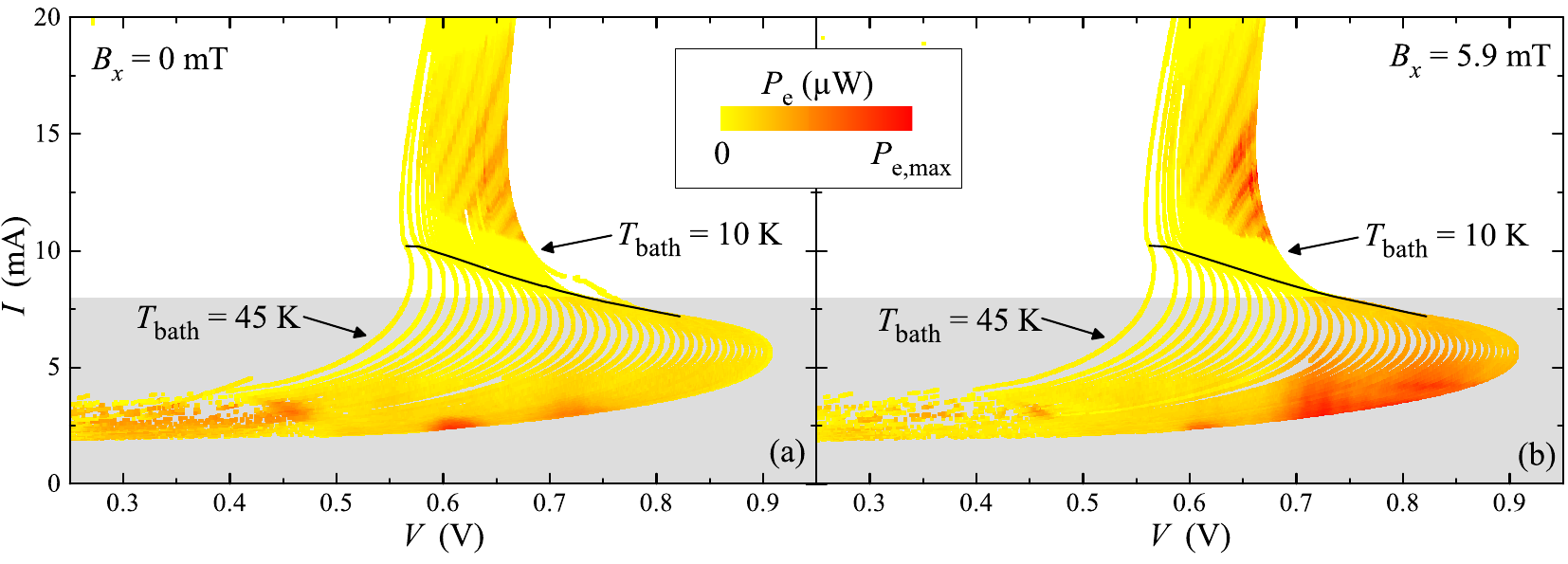}
\caption{Experimental data for a GBG structure: THz emission power $P_\mathrm{e}$ (color scale) for a large number of IVCs, measured at bath temperatures between 10\,K and 45\,K for (a) $B_x = 0$ and (b) $B_x = 5.9\,\mathrm{mT}$. In both (a) and (b) $P_\mathrm{e,max} = 27.5\,\mu\mathrm{W}$ for $I > 8\,\mathrm{mA}$, and $P_\mathrm{e,max} = 0.21\,\mu\mathrm{W}$ for $I< 8\,\mathrm{mA}$. Black lines in (a) and (b) indicate the $T_\mathrm{c}$ line.}
\label{fig:GBG_emi}
\end{figure*}
We also tested experimentally the potential benefit of a small magnetic field oriented along $x$, using a $75 \times 330\,\mu\mathrm{m}^2$ large GBG structure with $N \approx 760$, mounted on a sapphire lens.
Fig.~\ref{fig:GBG_emi} shows for (a) $B_x = 0$ and (b) $B_x = 5.9\,\mathrm{mT}$ ($0.32\,\Phi_0$ per junction) families of IVCs measured for $10\,\mathrm{K} \leq T_\mathrm{bath} \leq 45\,\mathrm{K}$. IVCs at 0 and $5.9\,\mathrm{mT}$ were measured alternately at given $T_\mathrm{bath}$. The accuracy in aligning the field with respect to out-of-plane tilts was better than $0.5^\circ$, and with respect to in-plane tilts it was about $2^\circ$. The simultaneously detected THz emission power $P_\mathrm{e}$, measured via a Ge bolometer, is plotted as a color scale. In the high-bias regime the maximum emission power $P_\mathrm{e,max}$ was $27.5\,\mu\mathrm{W}$, while at low bias it was $0.21\,\mu\mathrm{W}$. We thus use different values for $P_\mathrm{e,max}$ for $I > 8\,\mathrm{mA}$ and for $I < 8\,\mathrm{mA}$; for fixed $I$, $P_\mathrm{e,max}$ is the same in Fig.~\ref{fig:GBG_emi}(a) and (b). For $B_x = 0$ the emission is strong for $I$ between 10 and 20\,mA and $T_\mathrm{bath}$ between 10 and 40\,K. One notes short period oscillations in $P_\mathrm{e}$ which presumably are extrinsic in origin. These oscillations have been observed before\cite{Zhou15b,Rudau15,Tsujimoto16}. Apart from that the plots clearly show that, for $B_x = 5.9\,\mathrm{mT}$, over a wide range of currents and bath temperatures $P_\mathrm{e}$ has increased significantly, in some of the stripe-like regions up to a factor of 2.7. In the low-bias regime the effect is seen even more drastically, although on a much lower level of $P_\mathrm{e,max}$. The idea of applying a small field parallel to the long side of the stack, as suggested by the simulations, thus seems to work. For other field orientations the effect is not observed. Even a small field component perpendicular to the layers strongly suppresses $P_\mathrm{e}$\cite{Yamaki10}. For the measurements shown above for out-of-plane tilts larger than about $1^\circ$ (the precise value depends on bias current and bath temperature) the enhancement in emission power was lost in the high-bias regime. At low bias regime the critical tilts were on the order of $3-5^\circ$. Further, our simulations suggest that a field applied parallel to the short side is not helpful, because a $(0,n)$ mode with $n > 1$ is not promoted by an applied flux well below $\Phi_0/2$ per junction.
In Ref.~\onlinecite{Yamaki10} a 20$\%$ increase of $P_\mathrm{e}$ was observed for fields oriented in the $ab$ plane. Unfortunately the field direction relative to the mesa edges was not reported.  
\section{Summary}
\label{sec:summary}
In summary, we presented 3D simulations of the thermal and electromagnetic properties of a 
mesa consisting of 700 intrinsic junctions. Resonant modes can be excited in the stack both in the presence and in the absence of a hot spot, exhibiting standing waves either along the length ($(0,n)$ modes) or the width ($(1,0)$ mode) of the stack. Also more complex mixed modes were found. At fixed bath temperature and bias current, different modes can coexist. By applying a small magnetic field along the length of the stack it was possible to stabilize the $(1,0)$ mode, considered to be the best mode for THz emission. In experiment we found a strong field-induced enhancement of the emission power for a stand-alone stack for fields of around 5.9\,mT, small enough to be created by a simple electromagnet.

\acknowledgments
We gratefully acknowledge financial support by the National Natural Science Foundation of China (Grant Nos. 11234006 and 61501220), the Priority Academic Program Development of Jiangsu Higher Education Institutions, Jiangsu Provincial Natural Science Fund (BK20150561), the Deutsche Forschungsgemeinschaft (Project KL930/13-1), JSPS KAKENHI Grant Number 25289108, RFBR grants 14-02-91335 and 14-02-31374, and the EU-FP6-COST Action MP1201.


%
%
\bibliography{hot-spots-waves_v2}
\end{document}